\documentclass[12pt]{article}

\pdfoutput=1
\usepackage{amssymb, amsmath,amsfonts}
\usepackage{amscd}
\usepackage{latexsym,cite}
\usepackage{graphicx}
\usepackage{tikz}

\usepackage[pdftex, bookmarks=true,colorlinks=true,linkcolor=red,urlcolor=blue,citecolor=blue]{hyperref}
\input epsf

\usepackage{hyperref}
\hypersetup{
    colorlinks,
    citecolor=blue,
    filecolor=black,
    linkcolor=red,
    urlcolor=blue,
    linktoc=all
}

\usepackage[textheight=9in, textwidth=6.5in, letterpaper]{geometry}

\makeatletter\@addtoreset{equation}{section}\makeatother

\def\be{\begin{equation}}
\def\ee{\end{equation}}
\def\bea{\begin{eqnarray}}
\def\eea{\end{eqnarray}}
\newcommand{\nn}{\nonumber}

\makeatletter\@addtoreset{equation}{section}\makeatother

\newcommand{\preprint}[1]{\begin{table}[t]  
             \begin{flushright}               
             {#1}                             
             \end{flushright}                 
             \end{table}}                     
\renewcommand{\title}[1]{\vbox{\center\LARGE{#1}}\vspace{5mm}}
\renewcommand{\author}[1]{\vbox{\center#1}\vspace{5mm}}
\newcommand{\address}[1]{\vbox{\center\em#1}}

\begin{document}

\begin{titlepage}

\preprint{IFT-UAM/CSIC-16-032}
 
\begin{center}
\hfill \\
\hfill \\
\vskip 1cm

\title{
Odd viscosity in the quantum critical region of a holographic Weyl semimetal}
\vskip 0.5cm

{Karl Landsteiner}\footnote{Email: {\tt  karl.landsteiner@uam.es} }, 
 {Yan Liu}\footnote{Email: {\tt  yanliu.th@gmail.com} } and 
 {Ya-Wen Sun}\footnote{Email: {\tt  yawen.sun@csic.es} }

\address{Instituto de F\'\i sica Te\'orica UAM/CSIC, C/ Nicolas Cabrera 13-15,\\ 
Universidad Aut\'onoma de Madrid, Cantoblanco, 28049 Madrid, Spain}

\end{center}

\vskip 1.5cm

\abstract{ We study odd viscosity in a holographic model of a Weyl semimetal. The model is characterised by 
a quantum phase transition from a topological semimetal to a trivial semimetal state. 
Since the model is axisymmetric in three spatial dimensions there are two independent odd viscosities.  
Both odd viscosity coefficients are non-vanishing in the quantum critical region and non-zero only due to 
the mixed axial gravitational anomaly. It is therefore a novel example in which the mixed axial gravitational
anomaly gives rise to a transport coefficient at first order in derivatives at finite temperature. 
We also compute anisotropic shear viscosities and show that one of them violates the KSS bound. 
In the quantum critical region, the physics of viscosities as well as conductivities is governed by the quantum critical point. 
}


\end{titlepage}

{\em Introduction.--} One of the most surprising outcomes of string theory is the application of the AdS/CFT correspondence
to the physics of strongly interacting quantum many-body systems  
\cite{books}. 
The need to develop models that allow to address the question of real-time transport in strongly
interacting quantum fluids has arisen from experiments in completely different areas of physics: in the
quark gluon plasma generated in heavy ion collisions, the collective behavior of ultra-cold atoms, the strange metal phase of the high-$T_c$ superconductors and
most recently the hydrodynamic electronic flow observed in Graphene and similar materials \cite{JanZaanen,{sci1},{sci2},{sci3}}. 

Graphene is a ``Dirac'' semimetal in which the electrons are well described by the Dirac equation. The motion
of electrons in Graphene is however restricted to two spatial dimensions. In the last few years new materials
whose electronics is described by the Dirac or Weyl equation in three spatial dimensions have been demonstrated
\cite{WSMviewpoint,wsmreview2,{Hosur:2013kxa}}. These Weyl semimetals have a plethora of exciting and exotic transport properties related to the chiral anomaly of three dimensional relativistic fermions. 

As in Graphene the electron fluid within a Weyl semimetal might as well be strongly interacting due
to the smallness of the Fermi velocity compared to the speed of light. It seems therefore natural to ask if  
holography can be applied to such systems 
as well. In this case holography should play a similar important role for the understanding of quantum transport of Weyl semimetals as it already does in the theory of the quark gluon plasma \cite{CasalderreySolana:2011us}. 
In particular we ask the question if one can learn something new from holographic models utilizing universal properties of these materials such as the (effective) presence of chiral anomalies. We will address this question and answer it to the affirmative.

{\em Holographic Weyl semimetal.--} Recently a holographic model of a Weyl semimetal has been developed in \cite{{Landsteiner:2015lsa},Landsteiner:2015pdh}. Let us briefly review the most salient feature of this model. 
Its action is given by
\begin{align}
\label{eq:holomodel}
  S=&\int d^5x \sqrt{-g}\bigg[\frac{1}{2\kappa^2}\Big(R+\frac{12}{L^2}\Big)-\frac{1}{4e^2}\mathcal{F}^2-\frac{1}{4e^2}F^2  -(D_\mu\Phi)^*(D^\mu\Phi)-V(\Phi)  \\
 &+\epsilon^{\mu\nu\rho\sigma\tau}A_\mu\bigg(\frac{\alpha}{3} \Big(F_{\nu\rho} F_{\sigma\tau}+3 \mathcal{F}_{\nu\rho}  \mathcal{F}_{\sigma\tau}\Big)+\zeta  R^{\beta} _{~\delta\nu\rho}R^{\delta}_{~\beta\sigma\tau} \bigg)\bigg]\,,\nonumber
\end{align} 
with $\mathcal{F}_{\mu\nu}=\partial_\mu V_\nu-\partial_\nu V_\mu, F_{\mu\nu}=\partial_\mu A_\nu-\partial_\nu A_\mu$ and $D_\mu\Phi = (\partial_\mu - i q A_\mu)\Phi.$  
The holographic dictionary determines the field content of the model. The metric encodes the dynamics of the energy momentum tensor. There are two gauge fields. The first one, denoted by $V_\mu$, is dual to a conserved vector $U(1)$ current that can be identified with the electric current. The second one, $A_\mu$, is an axial gauge field. It couples to the 
complex scalar field $\Phi$ via an axial covariant derivative. The axial current suffers also from the axial anomaly which has three parts:
one is the electro-magnetic contribution to the axial anomaly, the second one is the purely axial $U(1)_A^3$ 
anomaly and the third one is the gravitational contribution to the axial anomaly (i.e. mixed axial gravitational anomaly).
These three anomalies are represented by the Chern-Simons terms in the action (\ref{eq:holomodel}).
The scalar field potential is chosen to be $V(\Phi) = m^2 |\Phi|^2 + \frac{\lambda}{2} |\Phi|^4$. The mass determines the dimension of the operator dual to $\Phi$ and we chose it to be
$m^2L^2 = -3$.\footnote{Here $L$ is the scale of the AdS space. In the following we set $2\kappa^2=e^2=L=1$.} 
The boundary value of the scalar field is dual to a mass deformation in the field theory. 

In \cite{Landsteiner:2015pdh} the boundary conditions\footnote{The same setup with different boundary conditions was also used in \cite{{Jimenez-Alba:2015awa},Sun:2016gpy} to realize axial charge dissipations in the study of negative magnetoresistivity of the holographic Dirac semimetal.}
\begin{equation}
 \lim_{r\rightarrow\infty} r\Phi = M\,,~~~~~~~ \lim_{r\rightarrow\infty} A_z = b
\end{equation}
together with asymptotic AdS behaviour of the metric were considered.
Choosing furthermore the scalar field charge $q=1$ and the scalar self coupling $\lambda=1/10$ it was
found that the model undergoes a quantum phase transition as function of the dimensionless parameter $M/b$. 
 Note that the mixed axial gravitational anomaly is included in the holographic Weyl semimetal model (\ref{eq:holomodel}) while it does not play any role in all the discussions of \cite{Landsteiner:2015pdh}, including the phase transition and electric conductivities. 
This model can be understood as a gravity analogue of the
Lorentz breaking Dirac system with Lagrangian 
\begin{equation}
 \left[ \gamma^\mu (i \partial_\mu - e v_\mu - \gamma_5 b \delta^z_\mu) + M\right] \Psi=0\,.
\end{equation}
This Lorentz breaking Dirac system has been used as a model for Weyl semimetals before in e.g. \cite{{burkov},Grushin:2012mt, {burkov1},Volovik:2016lwb}.

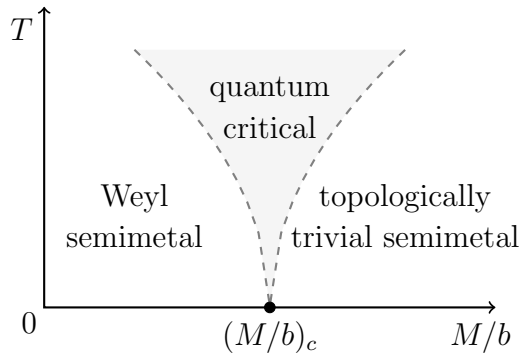
\begin{figure}[t]
\begin{center}
\begin{tikzpicture}
\draw [<->, thick] (0,4) -- (0,0) -- (6,0);
\draw[fill=white!20!white,draw=white] 
   (0.04,3.4) -- (0.04,0.06) -- (5.84,0.06)--(5.84,3.4)--(0.04,3.4);
\draw[fill=lightgray!15!white, draw=gray, dashed, thick,domain=1.2:4.8] plot (\x, {(4.6/1.8)*sqrt(abs(\x-3)))});
\draw node at (-0.2,-0.2){ $0$} node at (3,-0.4)  { $(M/b)_c$} node at (5.8,-0.4)  { $M/b$}
node at (-0.3,3.7)  { $T$};  
\draw node at (1.2,1.45){ Weyl}
node at (1.2,0.95){ semimetal}; 
\draw[black] node at (3,2.9){ quantum} node at (3,2.4){ critical}; 
\draw node at (4.8,1.45) { topologically}
node at (4.8,0.95) { trivial semimetal};
\draw [fill] (3,0) circle [radius=2pt];
\end{tikzpicture}
\caption{\small  The cartoon picture for the holographic Weyl semimetal model in the coupling constant ($M/b$) - temperature ($T$) plane.  At zero temperature a topological quantum phase transition occurs at the critical value $(M/b)_c$.  At finite temperature the dashed line is a smooth crossover and in the quantum critical regime the physics is governed by the quantum critical behaviour.} 
\label{fig-cartoon}
\end{center}
\end{figure}

At zero temperature for $M/b < 0.744$ the scalar field vanishes in the IR towards $r=0$ whereas the axial gauge field takes a non-vanishing value $A_z|_{r=0}=b_\mathrm{eff}$. In this regime the model has a non-vanishing anomalous Hall conductivity given by $\sigma_\text{AHE} = 8\alpha b_\mathrm{eff}$. 
For $M/b > 0.744$ the axial gauge field vanishes in the IR whereas the scalar field takes a finite value that is determined by the minimum of the potential $V'(\Phi)=0$.  
In this phase the anomalous Hall conductivity vanishes. The model undergoes therefore a topological quantum phase transition between a topological state of semimetal
state with non-vanishing anomalous Hall conductivity and a trivial semimetal state with vanishing anomalous Hall conductivity. There is an emergent Lifshitz symmetry at the critical point $M/b \simeq 0.744$ and it governs the quantum critical physics at finite temperature \cite{subir}.  Moreover, at low temperature the ohmic DC conductivity scales as $\sigma_{xx}=\sigma_{yy} = c T$ and $\sigma_{zz} = \tilde c T$ except near the quantum critical regime \cite{Landsteiner:2015pdh} as can be expected from a three dimensional Weyl- or Dirac semimetal. A cartoon illustration for our model (\ref{eq:holomodel}) is shown in Fig. \ref{fig-cartoon}.

{\em Viscosities.--} A necessary ingredient for the presence of odd viscosity is broken time reversal symmetry \cite{ASZ,Hoyos:2014pba, LandauLifschitz, {Haehl:2015pja}}. This is in principle provided by 
the axial gauge field background $b$. We note however that this is a UV parameter and we could expect that the viscosity is determined rather by the IR properties, similar to the anomalous Hall conductivity. It follows then
that in the topological trivial phase in which time reversal symmetry is restored at the endpoint of the holographic RG flow $b_\mathrm{eff} =0$ we should not expect
substantial odd viscosity. On the other hand one can also argue that odd viscosity should be absent in the topological phase at zero temperature. At weak coupling the argument
goes as follows: the low energy effective model describing a Weyl semimetal is 
\begin{equation}
 S = \int d^4x \bar\Psi( i \gamma^\mu \partial_\mu-e\gamma^\mu v_\mu - \gamma_5\gamma_z b_\mathrm{eff} ) \Psi\,.
\end{equation}
By a field redefinition the parameter $b_\mathrm{eff}$ can be removed from the action at the cost of introducing the anomalous effective term 
\begin{align}\label{eq:Sanom}
 \Gamma_\mathrm{anom} = \int d^4x\sqrt{-\gamma}(b_\mathrm{eff}\cdot z) \epsilon^{\mu\nu\rho\lambda}  \big(  
\alpha \mathcal{F}_{\mu\nu}\mathcal{F}_{\rho\lambda} +\frac{\alpha}{3} F_{\mu\nu}F_{\rho\lambda} + 
\zeta R^\alpha\,_{\beta \mu\nu} R^\beta\,_{\alpha \rho\lambda} \big) \,.
\end{align}
The anomaly (\ref{eq:Sanom}) encodes the response at zero temperature and shows that there is Hall conductivity but no odd (Hall) viscosity. Rather the gravitational response is third order in
derivatives as the Riemann curvature is second order in derivatives on the metric. 
We note that at finite temperature this derivative counting is not necessarily correct anymore. A well known example for this is the contribution of the mixed axial gravitational anomaly to the chiral vortical effect \cite{Landsteiner:2011cp, {Landsteiner:2011iq}}. As we will show now in our holographic model the gravitational contribution to the axial anomaly is also able to induce odd viscosity (a first order effect in derivatives) once temperature is switched on. 

In an axisymmetric system characterised by a time reversal breaking vector such as ${\vec b}$ there are 
seven\footnote{In a 3+1 dimensional axisymmetry system with a time reversal breaking vector, there are seven components in the viscosity tensor, including three shear viscosities, two odd viscosities and two bulk viscosities. Besides the four components below, there are another two bulk viscosities and one shear viscosity which come from the spin zero components of $xx+yy$ and $zz$, which we do not consider in this paper.} independent viscosities \cite{LandauLifschitz} in which there are 
two independent odd viscosity tensor components. 
We can define the viscosities via the Kubo formula
\be\eta_{ij,kl}=\lim_{\omega\to 0}\frac{1}{\omega} \text{Im}\big[G^R_{ij,kl}(\omega,0)\big]\,,
\ee
with the retarded Green's function of the energy momentum tensor
\be
G^R_{ij,kl}(\omega, 0)=-\int dt d^3x e^{i\omega t} \theta(t) \langle[T_{ij}(t,{\vec x}), T_{kl}(0, 0)]\rangle\,.
\ee
Since we chose our coordinates such that ${\vec b} = b \hat e_z$ is a convenient basis, for the two shear viscosities which are related to the symmetric part of the retarded Green's function under the exchange of $(ij)\leftrightarrow(kl)$
\be
\eta_\parallel=\eta_{xz,xz}=\eta_{yz,yz}\,,~~~~~~
\eta_\perp=\eta_{xy,xy}=\eta_{T,T}\\
\ee 
and for the two odd components of viscosity which are related to the antisymmetric part 
\be\label{eq:defoddvisoc}
\eta_{H_\parallel}=-\eta_{xz,yz}=\eta_{yz,xz}\,,~~~~\eta_{H_\perp}=\eta_{xy,T}=-\eta_{T,xy}
\ee
where $T$ denotes the index combination $xx-yy$. 
We note that $\eta_{H_\perp}$ can be understood as Hall viscosity in the plane orthogonal to $\vec b$ whereas
$\eta_{H_\parallel}$ is specific to axisymmetric three dimensional systems. The later has been shown to arise also via
the coupling of elastic gauge fields to the electron gas in Weyl semimetals \cite{Cortijo:2016yph}. In that case the odd or  Hall viscosity is best thought of as a property of the phonon gas arising via the electron-phonon Chern-Simons interactions. This effective Hall viscosity is related to the underlying Hall conductivity of the electron gas and arises from the electronic point of view as an axial Hall conductivity. In contrast here we will be dealing with Hall or odd viscosity as an intrinsic property of the strongly coupled electron fluid.  Hall viscosity arising from gravitational $\theta-$terms in holographic models dual to $2+1$ dimensional field theories has been studied before in e.g. \cite{Saremi:2011ab, {Chen:2012ti}, {Cai:2012mg}, {Zou:2013fua}, Liu:2014gto, {Fischler:2015kro},{Zhao:2015inu}}. In contrast our system is dual to a $3+1$ dimensional theory with a mixed  axial gravitational anomaly represented by the five dimensional gravitational Chern-Simons term in (\ref{eq:holomodel}). 

We use the following ansatz for the background at finite temperature
\begin{align}\label{eq:ansatzmetric}
 ds^2 &= -u dt^2 + \frac{dr^2}{u} + f(dx^2+dy^2) + h dz^2\nonumber\,\\
 A &= A_z dz~,~~~~ \Phi = \phi\,
\end{align}
with the fields $u, f, h, A_z, \phi$ real functions of $r$. Background equations of motion are summarised in appendix \ref{seca}. It turns out that they are independent of the 
Chern-Simons couplings $\alpha$ and $\zeta$ and the numerical solutions have been studied in \cite{Landsteiner:2015pdh}. At finite temperature there is a horizon at a finite value $r=r_0$. 
The entropy density is given by the area element of the horizon $s= 4\pi  f \sqrt{h} \big{|}_{r=r_0}$.
In order to probe the interesting non-trivial IR physics and relate our findings to possible applications to physical Weyl semimetals we should work at small temperatures. At higher temperatures it is rather the UV-completion of the model that is probed.  

\begin{figure}[h!]
\begin{center}
\includegraphics[width=0.47\textwidth]{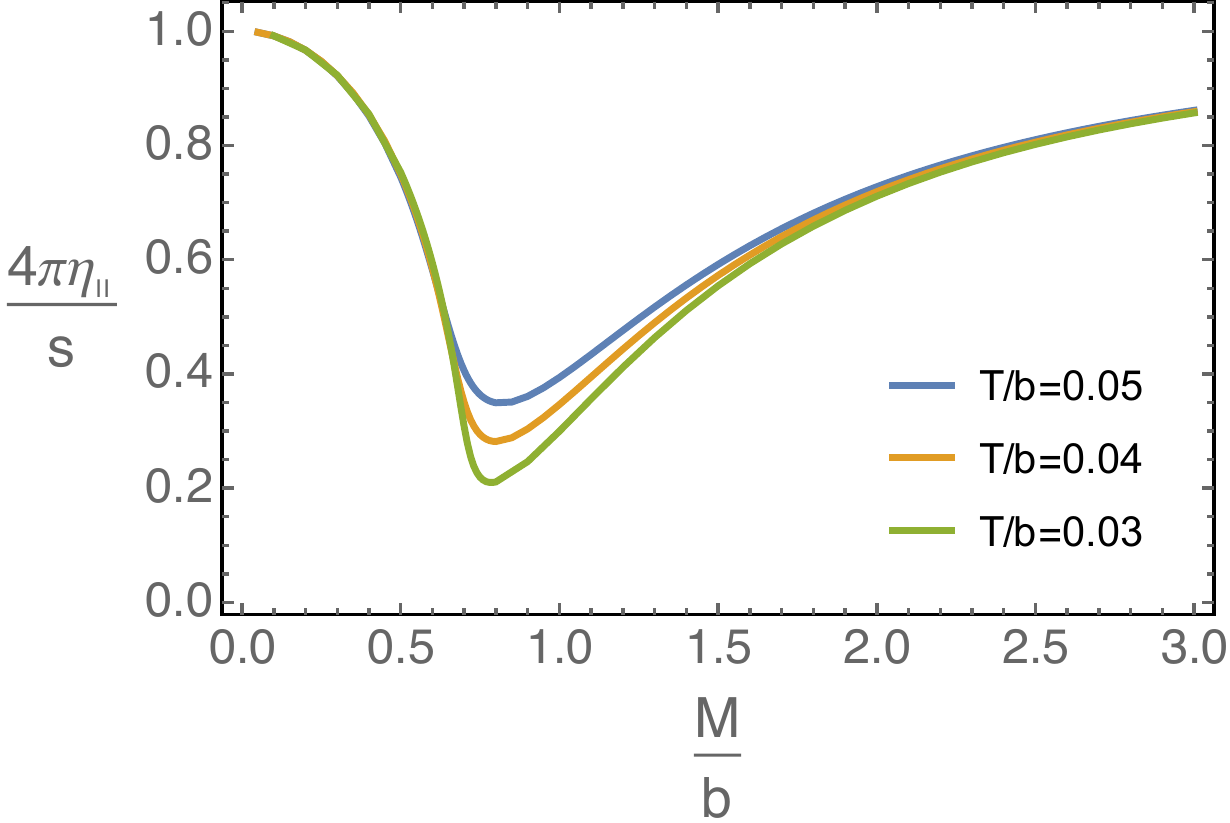}
\caption{\small The longitudinal shear viscosity over entropy density $4\pi \frac{\eta_\parallel}{s} $ as a function of $M/b$ at different temperatures.}
\label{fig:shearparallel}
\end{center}
\end{figure}

{\bf Longitudinal viscosity}: In order to compute the viscosities we switch on the following perturbations: 
$\delta g_{iz} = h_{iz} (r) e^{-i\omega t} $ , $ \delta A_i = a_i(r) e^{-i\omega t}$ for $i\in\{x,y\}$.
They form the complex combination $h_{\pm} = h_{xz} \pm i h_{yz}$ and $a_\pm = a_x\pm i a_y$.
The resulting equations of motion are rather cumbersome to treat, but after a lengthy but straightforward analysis the solutions to lowest order in $\omega$ can be written as (see appendix \ref{appb} for details)
\begin{align}
h_{\pm} = r^2-\frac{M^2}{3}+\frac{M^4 (2+3\lambda )}{18}\frac{\ln r}{ r^2} +
\frac{1}{r^2}\bigg[f_3+\frac{\omega}{4}\Big(\Big[i\frac{f^2}{\sqrt{h}}\pm 4\zeta \frac{q^2A_z\phi^2 f^2}{h}\Big]\Big{|}_{r=r_0}\Big)\bigg]+\dots\,
\end{align} 
near the conformal boundary.
Here $f_3 $ is the coefficient of the $1/r^2$ term in the asymptotic expansion of metric function (\ref{as-2}). 
From the first order term in $\omega$ we can read off the following two viscosity coefficients,
\begin{align}\label{eq:shearperp}
\text{dissipative~viscosity:}~~~&\eta_\parallel = \eta_{xz,xz} = \eta_{yz,yz} = \frac{f^2}{\sqrt{h}}\bigg{|}_{r=r_0}\\
\label{eq:oddvisperp}
\text{dissipationless~odd~viscosity:}~~~& \eta_{H_\parallel} = \eta_{yz,xz} = -\eta_{xz,yz}  = 4 \zeta \frac{q^2 A_z \phi^2 f^2}{h} \bigg{|}_{r=r_0}\,.
\end{align}
The dissipative viscosity is a form of shear viscosity and it is interesting to express it normalized to the entropy density $\frac{\eta_\parallel}{s}  = \frac{f}{4 \pi h}|_{r=r_0}$. As can be seen from Fig. \ref{fig:shearparallel} the shear viscosity drops significantly below the standard result of KSS bound \cite{Kovtun:2004de}. In view of the various results of 
violation of the KSS bound in anisotropic theories \cite{Rebhan:2011vd, {Jain:2014vka}} this is not  
unexpected. Still it is very interesting to note that the shear viscosity reaches a minimum in the quantum critical region of $M/b \approx 0.744$ as shown in Fig. \ref{fig:shearparallel}. 

\begin{figure}[h]
\begin{center}
\includegraphics[width=0.49\textwidth]{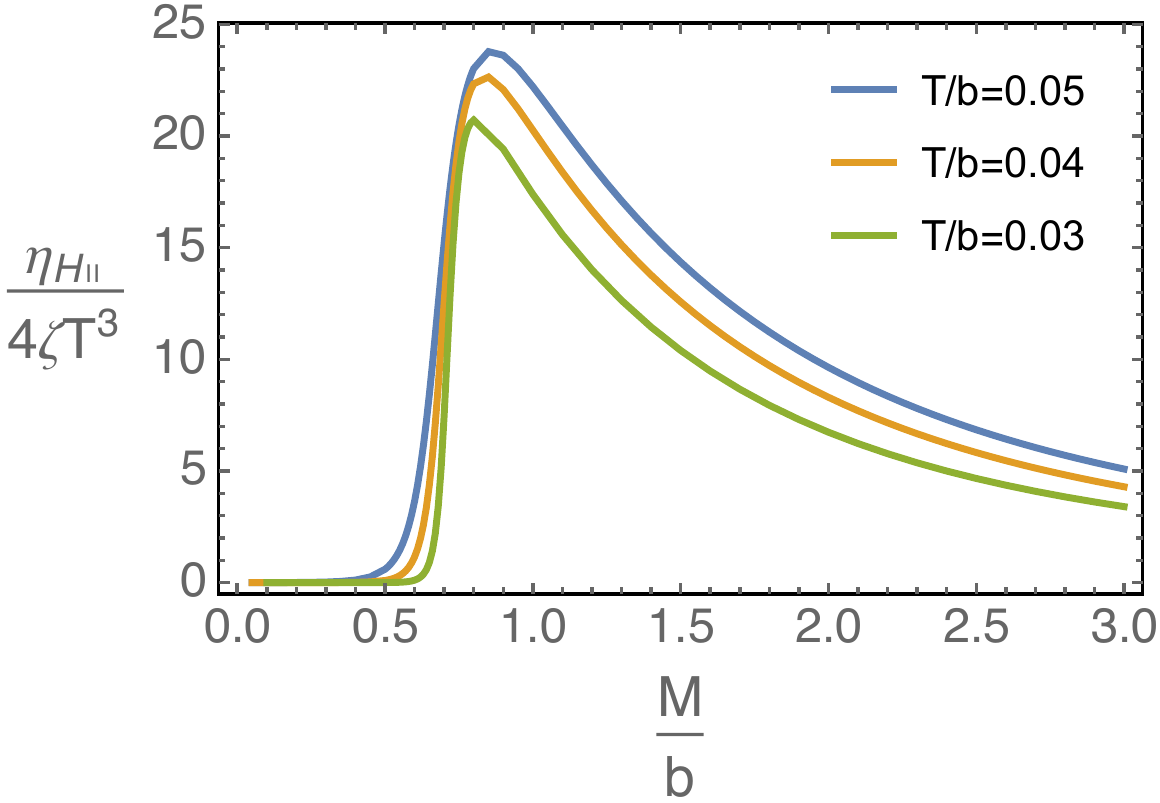}
\caption{\small Odd viscosity $\eta_{H_\parallel}$ as a function of $M/b$ at different temperatures.}
\label{fig:oddvisxzyz}
\end{center}
\end{figure}

The odd viscosity $\eta_{H_\parallel}$ as a function of $M/b$ for small but finite temperatures is shown in Fig.  \ref{fig:oddvisxzyz}. 
It is highly suppressed in the Weyl semimetal part of the phase diagram but rises steeply as the quantum critical region around
$M/b\simeq 0.744$ is entered. It peaks roughly at the critical value and then falls off in a somewhat slower fashion as $M/b$ increases.
The extreme $M/b\to\infty$ limit can be reached by setting $b=0$ keeping $M$ finite. In this case the field $A_z$ is simply zero along the holographic RG flow and from (\ref{eq:oddvisperp}) it follows that the odd viscosity vanishes again in this limit. 

\begin{figure}[h]
\begin{center}
\includegraphics[width=0.49\textwidth]{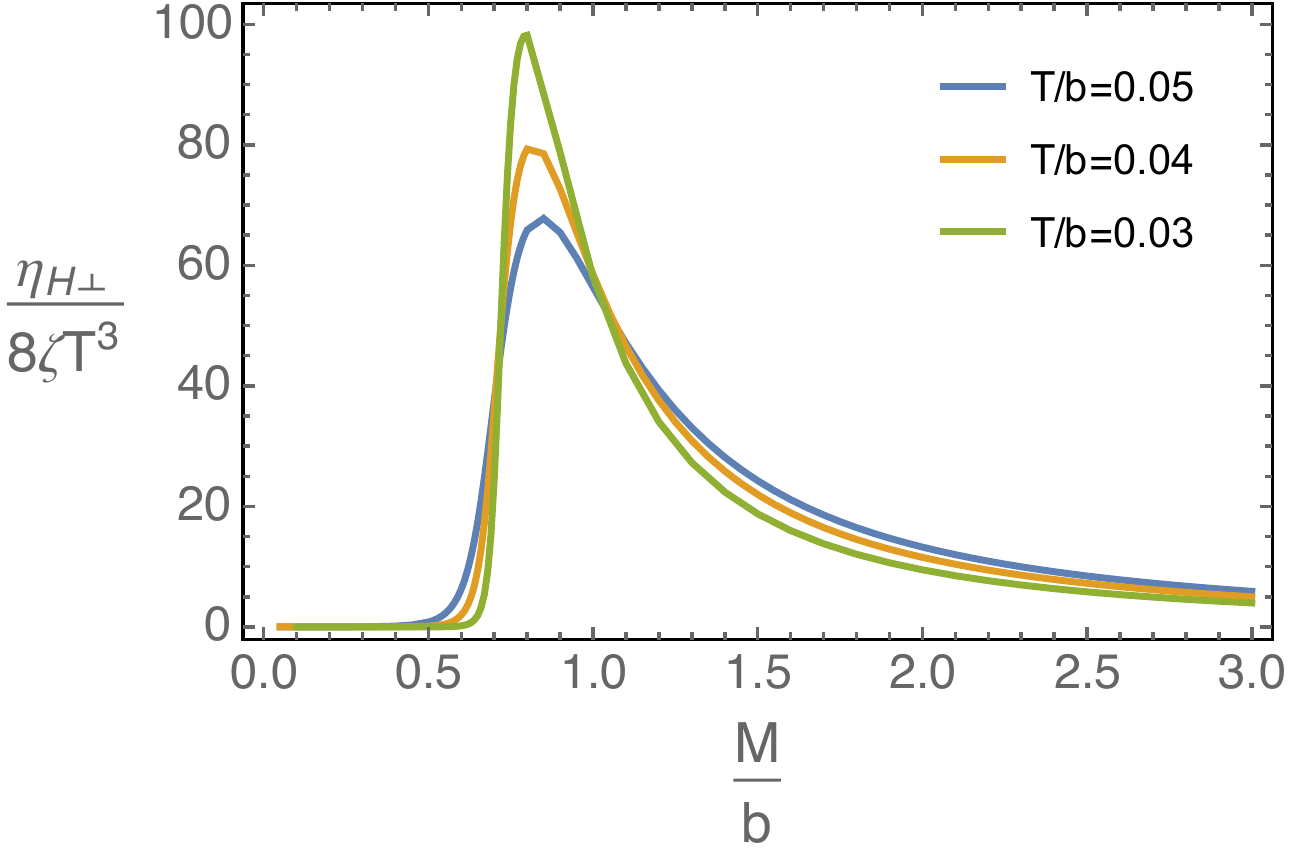}
\caption{\small Odd viscosity $\eta_{H_\perp}$ as a function of $M/b$ at different low temperatures.}
\label{fig:oddvis_xyL}
\end{center}
\end{figure}

{\bf Transverse viscosity}: 
We switch on the perturbations $\delta g_{xx} - \delta g_{yy} = 2 h_L(r) e^{- i \omega t}$ , $\delta g_{xy} = h_{xy}(r)  e^{- i \omega t}$
and form the complex combination $H_\pm = h_L \pm i h_{xy}$. Expanding to first order in $\omega$ we find (see appendix \ref{appc} for details)
\begin{align}
 H_\pm =  r^2-\frac{M^2}{3}+\frac{M^4(2+3\lambda)}{18}\frac{\ln r}{r^2}+ 
 \frac{1}{ r^2}\bigg(f_3+\frac{\omega}{4}\Big(\Big[ i f\sqrt{h}   \pm  8\zeta q^2\phi^2 fA_{z}\Big]\Big{|}_{r=r_0}\Big)\bigg)+\dots
\end{align} 
near the conformal boundary. Using the holographic dictionary we can read off the viscosities from the terms at first order in $\omega$ at order $1/r^2$ in the large $r$ expansion.
The dissipative viscosity is
\be
 \eta_\perp = f\sqrt{h} \Big{|}_{r=r_0}\,.
\ee
We note that in this case the KSS bound is exactly obeyed $\eta_\perp/s = 1/4\pi$.
The non-dissipative odd viscosity is 
\begin{equation}
 \eta_{H_\perp} =  8 \zeta q^2 \phi^2 f A_z \Big{|}_{r=r_0}\,.
\end{equation}
 Again we find that in the low temperature regime the non-dissipative odd viscosity has substantial support only in the quantum critical region around $M/b \simeq 0.744$ as can be seen from Fig. \ref{fig:oddvis_xyL}. 
On the other hand in the regime in which the horizon probes the UV completion of our holographic model such that $r_0  \rightarrow \infty$ (
the extreme high temperature regime)
we find $2\eta_{H_\parallel} =  \eta_{H_\perp} = 8 \zeta q^2 M^2 b$. In this high temperature regime there is no trace left of the quantum phase transition at zero temperature.

We also note that the analytic results on the viscosities (together with our previous result on the conductivities in \cite{Landsteiner:2015pdh}) allow us to obtain the non-trivial relation
\be
\frac{\eta_\parallel}{\eta_\perp} = \frac{2\eta_{H_\parallel}}{\eta_{H_\perp}}=\frac{\sigma_\parallel}{\sigma_\perp}=\frac{f}{h}\bigg{|}_{r=r_0} 
\ee
where $\sigma_\parallel=\sigma_{zz}=\frac{f}{\sqrt{h}}\big{|}_{r=r_0}, \sigma_\perp=\sigma_{xx}=\sigma_{yy}=\sqrt{h}\big{|}_{r=r_0}.$

\begin{figure}[h]
\begin{center}
\includegraphics[width=0.43\textwidth]{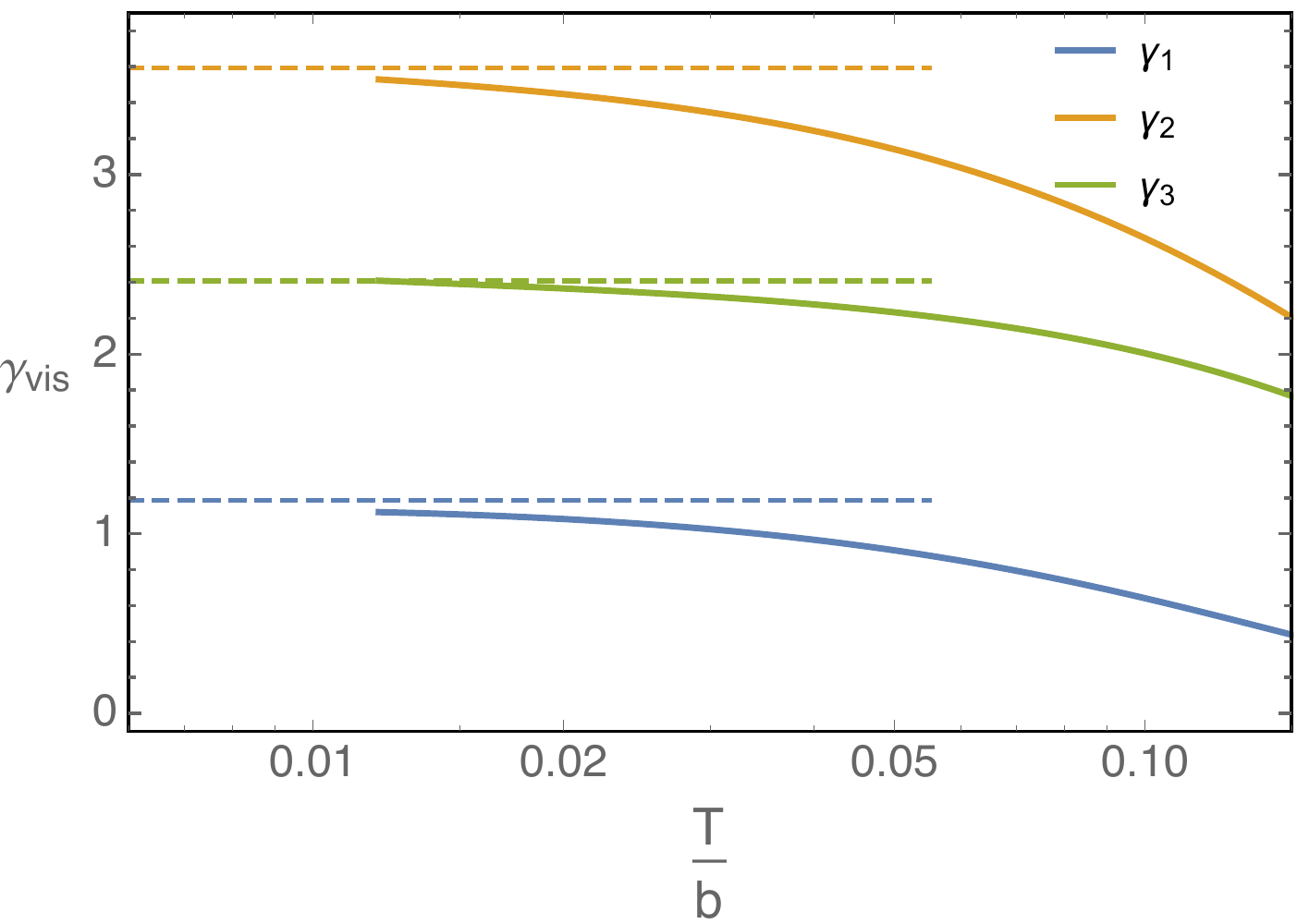}
\includegraphics[width=0.45\textwidth]{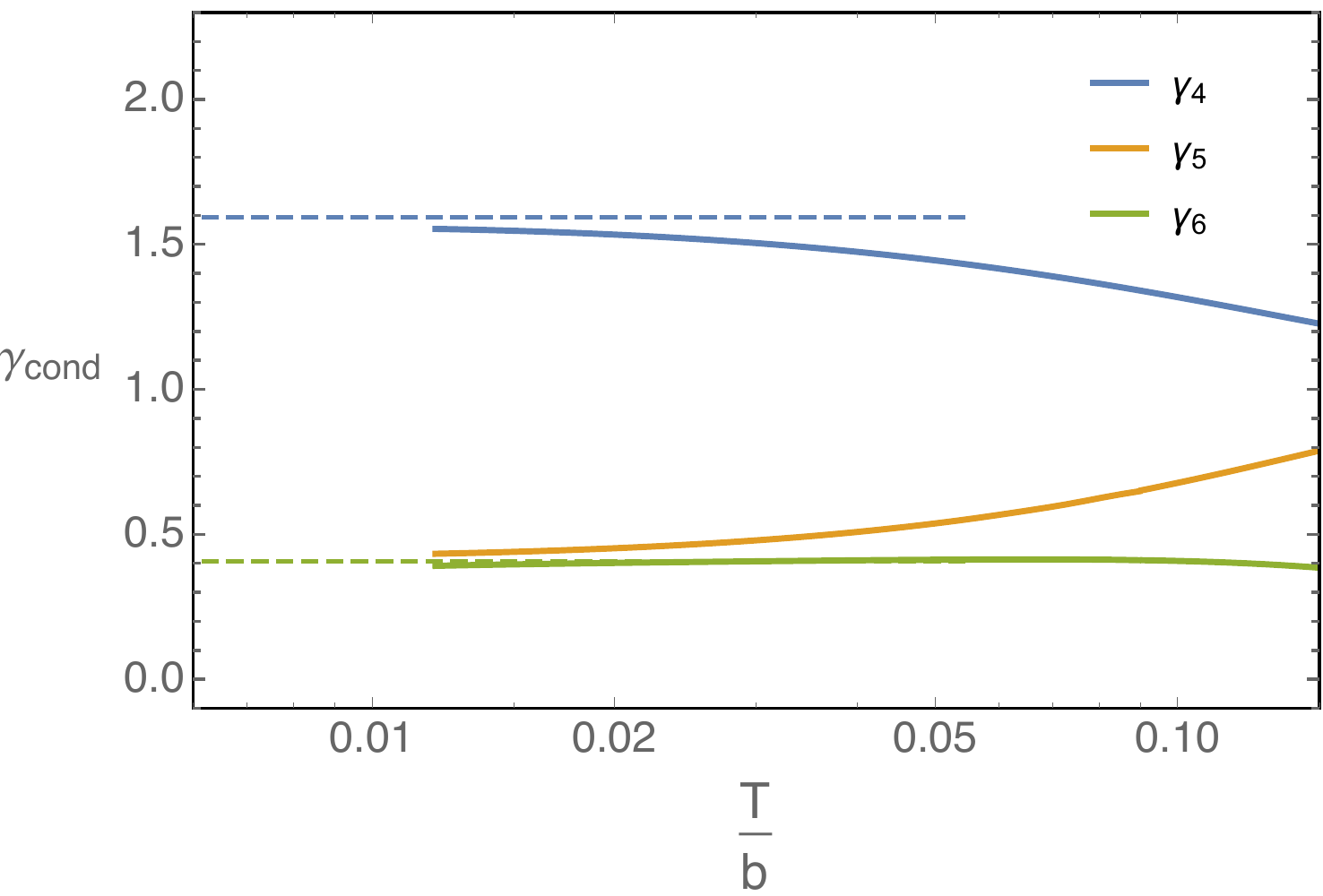}
\caption{\small The temperature scaling exponents $\gamma_i$ with $i\in\{1,\dots,6\}$  for viscosities $\eta_\parallel$, $\eta_{H_\parallel}$ and $\eta_{H_\perp}$ ({\em left}) and for electric conductivities $\sigma_\parallel$, $\sigma_\perp$ and $\sigma_\text{AHE}$ ({\em right}) at the critical value $M/b=0.744$ for low temperatures. The dashed lines in the picture denote the analytic values of the scaling exponents from the scaling analysis.}
\label{fig:exponent}
\end{center}
\end{figure}

Finally, let us comment on the temperature scaling behaviour of viscosities and conductivities in the quantum critical regime. At zero temperature, there is an emergent Lifshitz-like symmetry in the IR at the transition point $M/b\simeq 0.744$ and IR physics is invariant under $(t,x,y, r^{-1}) \to l (t,x,y, r^{-1}), z\to l^\beta z$ with the anisotropic scaling exponent $\beta\simeq 0.407$ together with  $f\to l^{-2} f, h\to l^{-2\beta} h, A_z\to l^{-\beta} A_z, \phi\to \phi$ \cite{Landsteiner:2015pdh}.  At very low temperature, since $T\to l^{-1} T$ we can obtain the temperature scaling behavior of the viscosities and conductivities near the critical region from scaling arguments.  More precisely, at the critical regime, when $M/b\to 0.744$, we have $\eta_\parallel/s\propto T^{\gamma_1}, \eta_{H_\parallel}\propto T^{\gamma_2},  \eta_{H_\perp}\propto T^{\gamma_3} $ with $(\gamma_1,\gamma_2, \gamma_3)=(2-2\beta, 4-\beta, 2+\beta)$\footnote{The shear viscosity bound proposed in \cite{Hartnoll:2016tri} is obeyed in our case.} and $\sigma_\parallel\propto T^{\gamma_4}, \sigma_\perp\propto T^{\gamma_5},  \sigma_\text{AHE}\propto T^{\gamma_6} $ with $(\gamma_4,\gamma_5, \gamma_6)=( 2-\beta, \beta, \beta)$ for low temperatures. In Fig. \ref{fig:exponent} we plot the temperature scaling exponents $\gamma_i$ with $i\in\{1,\dots,6\}$ of our numerical results at low temperatures at the critical value of $M/b$. 
We can see from the figure that the scaling exponents are approaching the analytic values when the temperature decreases.
These scaling dependences explain the peak/dip behavious of the transports of holographic Weyl semimetal in the critical regime. 

{\em Conclusion.--} 
Using the holographic model of \cite{Landsteiner:2015pdh} we have shown that substantial odd viscosities are present in the low temperature quantum critical region in which the  quantum crossover transition between the Weyl semimetal state and the trivial semimetal takes place. The odd viscosities are non-vanishing only if we add the gravitational Chern-Simons term in the holographic action which is dual to the mixed axial-gravitational anomaly. From the field theory perspective the odd viscosity is therefore a consequence of the presence of the gravitational contribution to the axial anomaly. We stress that after the chiral vortical effect this is the second example in which the naive derivative counting for anomaly induced transport coefficients breaks down. 

While our results have been obtained using a holographic model we do expect
that the lessons one can draw from it have much wider applicability. We strongly believe that the presence of odd viscosities is a universal feature of the quantum critical region of Weyl semimetal/semimetal transitions. 
At weak coupling this is the region which is characterised by quadratic band touching in one of the directions in momentum space (the $z$-direction of our model). Our holographic model gives therefore an exciting new prediction for the physics of the quantum critical region of Weyl semimetals.

\subsection*{Acknowledgments}
We thank R. G. Cai,  A. Cortijo, Y. Ferreiro,  F. D. M. Haldane, E. Megias, S. Sachdev, G. W. Semenoff, M.A.H. Vozmediano, Q. Yang, J. Zaanen for useful discussions. This work has been supported by project FPA2012-32828 and by the 
Centro de Excelencia Severo Ochoa Programme under grant SEV-2012-0249. 
The work of Y.W.S. was also supported by the European Union through a Marie Curie Individual Fellowship MSCA-IF-2014-659135.


\newpage
\appendix

\setcounter{equation}{0}
\setcounter{figure}{0}
\setcounter{table}{0}
\makeatletter

\section{Background equations of motion and fluctuations}
\label{seca}
\subsection{Background equations of motion}
The equations of motion from the action (\ref{eq:holomodel}) are
\bea
R_{\mu\nu}-\frac{1}{2}g_{\mu\nu}\Big(R-12 -\frac{\kappa^2}{2 e^2}(\mathcal{F}^2+F^2)-(D_\mu\Phi)^*(D^\mu\Phi)-V(\Phi)\Big)
-\frac{\kappa^2}{e^2} \big(\mathcal{F}_{\mu\rho}\mathcal{F}_{\nu}^{~\rho}~~~&& \nonumber \\+F_{\mu\rho}F_{\nu}^{~\rho}\big)
-4\kappa^2\zeta \epsilon_{\alpha\beta\rho\tau(\mu} \nabla_\delta (F^{\beta\alpha} R^{\delta~~\rho\tau}_{~\nu)})-\kappa^2 ((D_{\mu} \Phi)^* D_{\nu} \Phi+(D_{\nu} \Phi)^*D_{\mu} \Phi)&=&0\,,\nonumber\\
\nabla_\nu \mathcal{F}^{\nu\mu}+2\alpha\epsilon^{\mu\tau\beta\rho\sigma} F_{\tau\beta}\mathcal{F}_{\rho\sigma}&=&0\,,\nonumber\\
\nabla_\nu F^{\nu\mu}+\epsilon^{\mu\tau\beta\rho\sigma} \Big[\alpha\big(F_{\tau\beta}F_{\rho\sigma}
+\mathcal{F}_{\tau\beta}\mathcal{F}_{\rho\sigma}\big)+\zeta R^{\delta}_{~\xi\tau\beta}R^\xi_{~\delta\rho\sigma}\Big]+i q\big[\Phi (D^\mu\Phi)^*-\Phi^*(D^\mu\Phi)\big]&=&0\,,\nonumber\\
D_\mu D^\mu\Phi-m^2\Phi-\lambda\Phi^{*2}\Phi&=&0\,,\nonumber
\eea 
where $A_{(\alpha}B_{\beta)}=\frac{1}{2}(A_\alpha B_\beta+A_\beta B_\alpha).$

With ansatz (\ref{eq:ansatzmetric}) for the background, the equations of motion are the same as in \cite{Landsteiner:2015pdh} 
\bea
u''+\frac{h'}{2h}u'-\bigg(f''+\frac{f'h'}{2h}\bigg)\frac{u}{f}&=&0\,,\nonumber\\
\frac{f''}{f}+\frac{u''}{2u}-\frac{f'^2}{4f^2}+\frac{f'u'}{fu}-\frac{6}{u}+\frac{\phi^2}{2u}\Big(m^2+\frac{\lambda}{2}\phi^2
-\frac{q^2A_z^2}{h}\Big)-\frac{{A_z'}^2}{4h}+\frac{1}{2}\phi'^2&=&0\,,\nonumber\\
\frac{1}{2}{\phi'}^2+\frac{6}{u}-\frac{u'}{2u}\bigg(\frac{f'}{f}+\frac{h'}{2 h}\bigg)
-\frac{f'h'}{2fh}-\frac{f'^2}{4f^2}+\frac{1}{4h}{A_z'}^2
-\frac{\phi^2}{2u}\bigg(m^2+\frac{\lambda}{2}\phi^2+\frac{q^2A_z^2}{h}\bigg)&=&0\,,\nonumber\\
A_z''+\bigg(\frac{f'}{f}-\frac{h'}{2h}+\frac{u'}{u}\bigg)A_z'-\frac{2q^2\phi^2}{u}A_z&=&0\,,\nonumber\\
\phi''+\bigg(\frac{f'}{f}+\frac{h'}{2h}+\frac{u'}{u}\bigg)\phi'-\bigg(\frac{q^2{A_z}^2}{h}+m^2+\lambda\phi^2\bigg)\frac{\phi}{u}&=&0\,.\nonumber
\eea

The ansatz (\ref{eq:ansatzmetric}) is for finite temperature backgrounds in general. At zero temperature we can set $f=u$ and the five equations above will reduce to four equations. 



Near the conformal boundary, the background fields can be expanded as 
\bea\label{as-1}
u&=&r^2-\frac{M^2}{3}+\frac{M^4(2+3 \lambda)}{18}\frac{\ln r}{r^2} -\frac{ M_b}{3r^2}+\dots\,,\\
\label{as-2}
f&=& r^2-\frac{M^2}{3}+\frac{M^4(2+3 \lambda)}{18}\frac{\ln r}{r^2}+\frac{f_3}{r^2}\dots\,,\\
h&=& r^2-\frac{M^2}{3}+\Big(\frac{M^4(2+3 \lambda)}{18}+\frac{q^2b^2M^2}{2}\Big)\frac{\ln r}{r^2}+\frac{h_3}{r^2}+\dots\,,
\\
A_z&=&b-bM^2q^2\frac{\ln r}{r^2}+\frac{\xi}{r^2}+\dots\,,\\
\label{as-5}
\phi&=&\frac{M}{r}-\frac{\ln r}{6r^3}(2M^3+3b^2Mq^2+3M^3 \lambda)+\frac{O}{r^3}+\dots\,
\eea
with $h_3=\frac{1}{72}(-144f_3+14 M^4-72 MO+9 b^2M^2q^2+9M^4\lambda).$ 

We have two radially conserved quantities $(\sqrt{h}(u'f-uf'))'=0$ and $(u'\sqrt{h}f-\frac{h'}{\sqrt{h}}uf-A_zA_z'\frac{uf}{\sqrt{h}})'=0$, which give  $f_3=-\frac{1}{3}M_b+\frac{1}{4}Ts$ with $s$ the entropy density of the system in the unit $16\pi G=1$ and $2 b \xi-4 MO+b^2M^2q^2-3Ts+4 M_b+(\frac{7}{9}+\frac{\lambda}{2})M^4=0$ separately. Thus $h_3=-\frac{1}{3}M_b+\frac{1}{4}Ts-\frac{1}{2}b\xi-\frac{1}{8}b^2M^2q^2.$ 

The renormalised action is given by  
\bea\label{eq:on-shell}
S_{\text{ren}}&=&S+\int_{r=r_\infty}d^4 x \sqrt{-\gamma}\bigg(2K-6-|\Phi|^2-\frac{1}{2} R[\gamma]+\frac{1}{2}(\log r^2)\Big[\frac{1}{4}F^2+\frac{1}{4}\mathcal{F}^2~~\nonumber\\&&~~~~~+|D_m\Phi|^2+(\frac{1}{3}+\frac{\lambda}{2})|\Phi|^4-\frac{1}{4}\big( R^{ab}R_{ab}-\frac{1}{3}R^2\big)\Big]\bigg)
\eea
where $\gamma_{ab}$ is the induced metric on the boundary, $K$ is the extrinsic curvature and $R_{ab}[\gamma]$ is the 
intrinsic Ricci tensor.

\subsection{Classfication of fluctuations}

To study the transport properties of the system, we turn on the following perturbations and study the corresponding linearized equations of motion 
\be \delta g_{\mu\nu}=h_{\mu\nu} e^{-i\omega t}, ~~~\delta V_\mu= v_\mu e^{-i\omega t}, ~~~\delta A_\mu= a_\mu e^{-i\omega t},~~~\delta \Phi= (\phi_1+i\phi_2) e^{-i\omega t}.\ee
Since the time-reversal symmetry breaking parameter $b$ is along the $z$ direction, these perturbations can be classified according to their spin under the rotation symmetry group $SO(2)$ in the $xy$ plane as follows.\footnote{We chose the radial gauge $a_r=v_r=g_{\mu r}=g_{r\mu}=0.$}  
\begin{itemize}

\item Spin 2: $ h_{xy}, h_T=\frac{1}{2}\big(h_{xx}- h_{yy}\big)$. These two fields couple together and they are responsible for the transverse shear and odd viscosities (appendix \ref{appc}). 

\item Spin 1: $h_{xz}, h_{yz}, a_x, a_y, v_x, v_y, h_{tx}, h_{ty}$. The equations for $v_x,~v_y$ couple together and decouple from other modes. The mixed anomaly term plays no role in the euqations of $v_x,~v_y$, thus the transverse electric conductivities are the same as in \cite{Landsteiner:2015pdh}. 
The modes $h_{xz}, h_{yz}, a_x, a_y$ couple with each other and they contribute to the longitudinal shear and odd viscosities (appendix \ref{appb}). The modes $h_{tx}, h_{ty}$ couple together which are responsible for the thermal conductivity (appendix \ref{appd}). 

\item Spin 0: $h_{tz}, h_{zz}, h_{xx}+h_{yy}, a_t, a_z, v_t, v_z,  \phi_1, \phi_2$. The equations for $v_z$ is still the same as in \cite{Landsteiner:2015pdh}. This sector will also contribute to one shear viscosity and two bulk viscosities. For our purpose of studying the odd viscosity we do not study these modes in this paper.

\end{itemize}

The equations of motion for the fluctuations $v_x, v_y$ and $v_z$ are the same as the case without mixed anomaly term \cite{Landsteiner:2015pdh}. Given the fact that the background is also the same as in \cite{Landsteiner:2015pdh}, the discussions on the phase transitions and electric conductivities in \cite{Landsteiner:2015pdh} still apply straightforwardly and the mixed axial gravitational anomaly does not play any role here. 

\section{Longitudinal viscosity}
\label{appb}
To compute the longitudinal viscocities of this system, we perturb the background solutions by  
$\delta g_{xz}=h_{xz}(r) e^{-i\omega t}\,, \delta g_{yz}=h_{yz}(r) e^{-i\omega t}\,, 
\delta A_{x}=a_{x}(r) e^{-i\omega t}\,, \delta A_{y}=a_{y}(r) e^{-i\omega t}\,.$
After redefining the fields 
\be \label{eq-tv-red1}
Y_\pm=g^{xx}\big(h_{xz}\pm i h_{yz}\big)=\frac{1}{f}\big(h_{xz}\pm i h_{yz}\big)\,,~~~ a_\pm=a_x\pm i a_y\,,
\ee 
we have the following linearized equations 
\bea\label{eomtv}
\Big(1\pm\frac{2\zeta\omega}{\sqrt{h}}C_1\Big)Y_{\pm}''+\Big[\frac{2f'}{f}+P_1\pm\frac{2\zeta\omega}{\sqrt{h}}\Big(\frac{2f'}{f}C_1+D_1\Big)\Big]Y_{\pm}'+\Big[\frac{\omega^2}{u^2}~~~~~~~~~~~~~~~~~~~~~~~\nonumber\\ \pm\frac{2\zeta\omega}{\sqrt{h}}\big(E_1+\frac{f'}{f}D_1+\frac{f''}{f}C_1\big)\Big]Y_{\pm}+\Big[\frac{A_z'}{f}\pm\frac{2\zeta\omega}{\sqrt{h}}\frac{S_1}{f} \Big]a_\pm' +\Big[\frac{2q^2A_z\phi^2}{uf}\pm
\frac{2\zeta\omega}{\sqrt{h}}\frac{W_1}{f}\Big]a_\pm&=&0\,,~~\nonumber\\
a_{\pm}''+ \bigg(\frac{u'}{u}+\frac{h'}{2h}\bigg)a_{\pm}'+\bigg(\frac{\omega^2}{u^2}-\frac{2q^2\phi^2}{u}\pm\frac{8\alpha \omega A_z'}{u\sqrt{h}}\bigg)a_{\pm}~~~~~~~~~~~~~~~~~~~~~~~~~~~~~~~~~~~~\nonumber \\-\frac{f}{h}\Big[A_z'\pm\frac{2\zeta\omega}{\sqrt{h}}S_1\Big]Y_{\pm}'\pm\frac{2\zeta \omega}{\sqrt{h}}\Big(-\frac{f' S_1}{h} +f G_1\Big)Y_{\pm}&=&0\,,~~~~
\eea
where the coefficients $C_1, P_1, D_1, E_1,  S_1, W_1, G_1$ are 
\bea\label{app-co1}
C_1(r)&=&2 A_z'\,,\nonumber\\
P_1(r)&=&\frac{u'}{u}-\frac{h'}{2h}\,,\nonumber\\
D_1(r)&=& 2A_z''+2 \Big(\frac{u'}{u}-\frac{h'}{h}\Big)A_z'\,,\nonumber\\
E_1(r)&=&-\Big(\frac{u'}{u}+\frac{f'}{f}\Big)A_z''+\Big(\frac{2\omega^2}{u^2}+\frac{f'^2}{f^2}+\frac{f'h'}{fh}-\frac{3f'u'}{fu}+\frac{h'u'}{hu}-\frac{f''}{f}-\frac{u''}{u}\Big)A_z'\,,\nonumber\\
S_1(r)&=& 2 h''-\frac{h'^2}{h}-\frac{2h f''}{f}+\frac{hf'^2}{f^2}\,,\\
W_1(r)&=&-\frac{2hf'^3}{f^3}+\frac{2h'^3}{h^2}+\frac{2hf'^2u'}{f^2u}-\frac{2h'^2u'}{hu}+\frac{4hf'f''}{f^2}-\frac{f''h'}{f}-\frac{3hf''u'}{fu}+\frac{f'h''}{f}\nonumber\\&&~~~~-\frac{4h'h''}{h}+\frac{3u'h''}{u}-\frac{hu''f'}{uf}+\frac{u''h'}{u}-\frac{2hf'''}{f}+2h'''\,,\nonumber\\
F_1(r)&=&-S_1/h=-\frac{2h''}{h}+\frac{2f''}{f}-\frac{f'^2}{f^2}+\frac{h'^2}{h^2}\,,\nonumber\\
G_1(r)&=&\frac{u'f'^2}{uf^2}-\frac{u'h'^2}{uh^2}-\frac{f''h'}{fh}-\frac{u'f''}{uf}+\frac{f'h''}{fh}+\frac{u'h''}{uh}-\frac{u''f'}{uf}+\frac{u''h'}{uh}\,.\nonumber
\eea

\subsection{Finite temeprature solutions}
\label{secb1}
To calculate the retarded Green's function, we take the following redefination for the perturbations at finite temperature and small $\omega/T$
\be \label{eq:expand1}
Y_\pm =u^{-i\omega/(4\pi T)} \big(Y_\pm^{(0)}+\omega Y_\pm^{(1)}+\dots\big)\,,~~~ 
 a_\pm=u^{-i\omega/(4\pi T)} \big(a_\pm^{(0)}+\omega a_\pm^{(1)}+\dots\big)
 \ee  
 where $Y_\pm^{(0, 1)}, a_\pm^{(0,1)}$ are regular functions of $r-r_0$ near the horizon. Then the equations for these fields at zeroth order in $\omega$ become 
\bea\label{eom-tranvis-eq1}
\Big(\frac{uf^2}{\sqrt{h}}Y_\pm^{(0)'}+\frac{uf}{\sqrt{h}}A_z'a_\pm^{(0)}\Big)'&=&0\,,\\
a_\pm^{(0)''}+\Big(\frac{u'}{u}+\frac{h'}{2h}\Big)a_\pm^{(0)'}-\frac{2q^2\phi^2}{u}a_\pm^{(0)}-\frac{f A_z'}{h}Y_\pm^{(0)'}&=&0\,.
\eea
At first order in $\omega$ we have\footnote{The $\omega^2$ term in $E_1$ of (\ref{app-co1}) should be ignored in the following equations.} 
\bea
Y_\pm^{(1)''}+\Big(\frac{u'}{u}+\frac{2f'}{f}-\frac{h'}{2h}\Big)Y_\pm^{(1)'}+\frac{A_z'}{f}a_\pm^{(1)'}+\frac{2q^2\phi^2A_z}{uf}a_\pm^{(1)}\pm\frac{2\zeta}{\sqrt{h}f}S_1a_\pm^{(0)'}
~~~~~~&&\nonumber\\ +\Big[-\frac{iA_z'u'}{4\pi Tfu} 
\pm\frac{2\zeta}{\sqrt{h}f}
W_1\Big]a_\pm^{(0)}+\frac{2\zeta}{\sqrt{h}}C_1Y_\pm^{(0)''}+\Big[-\frac{iu'}{2\pi Tu}\pm\frac{2\zeta }{\sqrt{h}} \Big(D_1+\frac{2f'}{f}C_1\Big) \Big]Y_\pm^{(0)'}~~~~~~&&\nonumber\\
+\Big[-\frac{i}{4\pi T}\Big(\frac{u''}{u}+\frac{2f'u'}{f u}-\frac{u'h'}{2 uh}\Big)
+\frac{2\zeta}{\sqrt{h}}\Big(E_1+\frac{f'}{f}D_1+\frac{f''}{f}C_1\Big)\Big]Y_\pm^{(0)}&=&0\,,\nonumber\\
a_\pm^{(1)''}+\Big(\frac{u'}{u}+\frac{h'}{2h}\Big)a_\pm^{(1)'}-\frac{2q^2\phi^2}{u}a_\pm^{(1)}-\frac{f A_z'}{h}Y_\pm^{(1)'}\mp\frac{2\zeta}{\sqrt{h}}\frac{fS_1}{h} Y_\pm^{(0)'}
+\Big[\frac{ifA_z'u'}{4\pi Thu}~~~~~~&&\nonumber\\
\pm\frac{2\zeta}{\sqrt{h}}\big(-\frac{f'S_1}{h}+fG_1\big)\Big]Y_\pm^{(0)}-\frac{iu'}{2\pi Tu}a_\pm^{(0)'}+
\Big[\pm\frac{8\alpha}{u\sqrt{h}}A_z'-\frac{i}{4\pi T}\Big(\frac{u''}{u}+\frac{u'h'}{2uh}\Big)
\Big]a_\pm^{(0)}&=&0\,.\nonumber
\eea

The first zeroth order equation (\ref{eom-tranvis-eq1}) reduces to $fY_\pm^{(0)'}+A_z'a_\pm^{(0)}=0$ under regular boundary conditions for the fields $Y_\pm^{(0)}$ and $a_\pm^{(0)}$. We have two possible classes of zeroth order solutions due to two remaining integration constants after imposing the infalling boundary conditions at the horizon. We focus on the zeroth order solutions $Y_{\pm}^{(0)}=1, ~a_\pm^{(0)}=0$, i.e. the $a_\pm$ modes are sourceless at leading order\footnote{The second regular solution at  zeroth order is $a^{(0)}_{\pm}$ being the solution of 
$a_\pm^{(0)''}+\big(\frac{u'}{u}+\frac{h'}{2h}\big)a_\pm^{(0)'}-\big(\frac{2q^2\phi^2}{u}-\frac{A_z'^2}{h}\big)a_\pm^{(0)}=0 $
and $Y^{(0)}_{\pm}=c-\int_{r_0}^r\frac{A_z'}{f}a_\pm^{(0)}d\tilde{r}.$ By choosing suitable $c$ we can set $Y_\pm^{(0)}$ sourceless near the boundary.}. With regular boundary condition at the horizon and sourceless boundary condition at the boundary $Y_{\pm}^{(1)}$ can be solved to be\footnote{\label{foot11}We have used the fact that the near horizon expansion of the background equation of motion gives us that $A_{z2}=\frac{q^2\phi_1^2}{2\pi T}A_{z1}$ with $A_{z2}, \phi_1, A_{z1}$ the horizon values of $A_z'$, $\phi$ and $A_z$.} 
\be\label{eq:solY1} 
Y_{\pm}^{(1)}=\frac{i}{4\pi T}\ln u+\int_{r_0}^r\Big[-\frac{A_z'a_\pm^{(1)}}{f} +\Big(-i\frac{f_1^2}{\sqrt{h_1}}\mp
4\zeta \frac{q^2 A_{z1}\phi_1^2f_1^2}{h_1}\Big)\frac{\sqrt{h}}{uf^2} \pm 2\zeta \Big(\frac{u}{f}\Big)'\frac{f}{u\sqrt{h}}A_z'\Big]d\tilde{r}\,,
\ee 
where $f_1, h_1, A_{z1}$ are near horizon values of $f, h, A_z$ respectively and $a_{\pm}^{(1)}$ is determined by 
\bea a_\pm^{(1)''}+\Big(\frac{u'}{u}+\frac{h'}{2h}\Big)a_\pm^{(1)'}-\Big(\frac{2q^2\phi^2}{u}-\frac{A_z'^2}{h}\Big)a_\pm^{(1)}+\Big(i\frac{f_1^2}{\sqrt{h_1}}\pm
4\zeta \frac{q^2 A_{z1}\phi_1^2f_1^2}{h_1}\Big)\frac{A_z'}{uf\sqrt{h}}~~~&&\nonumber\\
\pm \frac{2\zeta}{\sqrt{h}}\Big[-
\Big(\frac{u}{f}\Big)'\frac{f^2}{uh}{A_z'}^2-f'\frac{S_1}{h}+f G_1
\Big]&=&0\,.\nonumber
\eea
It follows that near the conformal boundary one specific solution of $a_{\pm}^{(1)}$ behaves as 
\be
a_\pm^{(1)}=a_\pm^{(s0)}-M^2q^2a_\pm^{(s0)} \frac{\ln r}{r^2}+\frac{a_\pm^{(2)}}{2r^2}+\cdots\,.
\ee
After adding the homogeneous solution for $a_\pm^{(1)}$ with infalling boundary condition near the horizon to the specific solution, one can always choose the combined solution of $a_\pm^{(1)}$ such that the source term in $a_\pm^{(1)}$ is zero. From (\ref{eq-tv-red1}), (\ref{eq:expand1}) and (\ref{eq:solY1}) we have the following solution at the conformal boundary\footnote{Note that we have set the source term to be 1 and one can multiply the solution with an arbitrary constant to get a new solution.} 
\bea
h_{xz}\pm i h_{yz}&=&fu^{-\frac{i\omega}{4\pi T}}\Big(1+\omega Y_\pm^{(1)}+\dots\Big)
=f \Big(1+\frac{\omega}{4r^4}\Big(i\frac{f_1^2}{\sqrt{h_1}}\pm
4\zeta \frac{q^2 A_{z1}\phi_1^2f_1^2}{h_1}\Big)+\dots\Big)
\nn\\
&=& r^2-\frac{M^2}{3}+\frac{M^4 (2+3\lambda )}{18}\frac{\ln r}{ r^2}
+\frac{1}{r^2}\bigg[f_3+\frac{\omega}{4}\Big(i\frac{f_1^2}{\sqrt{h_1}}\pm
4\zeta \frac{q^2 A_{z1}\phi_1^2f_1^2}{h_1}\Big)\bigg]+\dots~~~~~
\eea
From this solution we can get the following conformal boundary solutions for $h_{xz}$ and $h_{yz}$ up to first order in $\omega$ 
\be
h_{xz}= r^2-\frac{M^2}{3}+\frac{M^4 (2+3\lambda )}{18}\frac{\ln r}{ r^2}+\frac{1}{r^2}\bigg(f_3+\frac{i\omega}{4}\frac{f_1^2}{\sqrt{h_1}}\bigg)+\dots\,, ~~
h_{yz}= -\frac{i\omega}{4r^2}\bigg(
4\zeta \frac{q^2 A_{z1}\phi_1^2f_1^2}{h_1}\bigg)+\dots\nn,
\ee where $h_{yz}$ is sourceless. Similarly due to the rotation symmetry in the $x$-$y$ plane we can also obtain the following solutions with sourceless $h_{xz}$ by multiplying the solutions of $Y_{\pm}$ by $\pm i$ 
\be
h_{xz}=\frac{i\omega}{4r^2}\bigg(
4\zeta \frac{q^2 A_{z1}\phi_1^2f_1^2}{h_1}\bigg)+\dots\,,~~~
h_{yz}= r^2-\frac{M^2}{3}+\frac{M^4 (2+3\lambda )}{18}\frac{\ln r}{ r^2}+\frac{1}{r^2}\bigg(f_3+\frac{i\omega}{4}\frac{f_1^2}{\sqrt{h_1}}\bigg)+\dots\,. \nn
\ee 
\subsection{Holographic renormalization}
Near the conformal boundary, the fluctuations responsible for the longitudinal viscosities are 
\bea
h_{xz}&\simeq&h_{xz}^{(0)}r^2+h_{xz}^{(0)}\big(-\frac{M^2}{3}+\frac{\omega^2}{4}\big)+
\frac{\ln r}{r^2}\Big[\frac{h_{xz}^{(0)}}{144}(16 M^4+24 M^4\lambda-6 M^2\omega^2+9\omega^4)\nn\\
&&~~~~+72 a_x^{(0)}b M^2 q^2\Big]+\frac{h_{xz}^{(2)}}{4r^2}+\dots\,, \nn\\
h_{yz}&\simeq&h_{yz}^{(0)}r^2+h_{yz}^{(0)}\big(-\frac{M^2}{3}+\frac{\omega^2}{4}\big)+
\frac{\ln r}{r^2}\Big[\frac{h_{yz}^{(0)}}{144}(16 M^4+24 M^4\lambda-6 M^2\omega^2+9\omega^4)\nn\\
&&~~~~+72 a_y^{(0)}b M^2 q^2\Big]+\frac{h_{yz}^{(2)}}{4r^2}+\dots\,, \nn\\
a_{x}&\simeq&a_x^{(0)}+a_x^{(0)}\big(-M^2 q^2+\frac{\omega^2}{2}\big)\frac{\ln r}{r^2}+\frac{a_x^{(2)}}{r^2}+\dots \,,\nn\\
a_{y}&\simeq&a_y^{(0)}+a_y^{(0)}\big(-M^2 q^2+\frac{\omega^2}{2}\big)\frac{\ln r}{r^2}+\frac{a_y^{(2)}}{r^2}+\dots\,.\nn
\eea

Up to the quadratic order in perturbations, from (\ref{eq:on-shell}) we have the following renormalized on shell action\footnote{Since we focus on the viscocities, i.e. the retarded Green's function in the hydrodynamic limit, we only write out the result up to the leading order in the frequency. 
}
\bea
S_{\text{on-shell}}&=&
\int \frac{d\omega}{2\pi} d^3x \bigg[ a_x^{(0)}(-\omega)a_x^{(2)}(\omega)+a_y^{(0)}(-\omega)a_y^{(2)}(\omega)+\frac{8}{3}i\alpha \omega ba_y^{(0)}(-\omega)a_x^{(0)}(\omega)
\nonumber\\&&~~~~
-\frac{8}{3}i\alpha \omega b a_x^{(0)}(-\omega)a_y^{(0)}(\omega)
+h_{xz}^{(0)}(-\omega)h_{xz}^{(2)}(\omega)+h_{yz}^{(0)}(-\omega)h_{yz}^{(2)}(\omega)
\nonumber\\&&~~~~
+\mathcal{O}(\omega^2)+\text{contact terms}\bigg]\nonumber,
\eea
where 
\bea
\text{contact terms}&=& (-2\zeta-bM^2 q^2) a_x^{(0)}(-\omega)h_{xz}^{(0)}(\omega)-\frac{1}{2}bM^2q^2 h_{xz}^{(0)}(-\omega)a_x^{(0)}(\omega)\nonumber\\~~~
&&+(-2\zeta-bM^2 q^2) a_y^{(0)}(-\omega)h_{yz}^{(0)}(\omega)-\frac{1}{2}bM^2q^2 h_{yz}^{(0)}(-\omega)a_y^{(0)}(\omega)\nonumber\\~~~&&
+M^2 q^2\big(a_y^{(0)}(-\omega)a_y^{(0)}(\omega)+a_x^{(0)}(-\omega)a_x^{(0)}(\omega)\big)
+\big(h_{xz}^{(0)}(-\omega)h_{xz}^{(0)}(\omega)
\nonumber\\~~~&&
+h_{yz}^{(0)}(-\omega)h_{yz}^{(0)}(\omega)\big)\Big(4 f_3-\frac{7M^4}{12}+2MO-\frac{M_b}{3}-\frac{M^4\lambda}{2}\Big)+\mathcal{O}(\omega^2)\nn.
\eea
Note that the contact terms are real and it will not contribute to the imaginary part of the retarded Green's function. Thus if we normalize the source terms to be 1, up to the first order in $\omega$ we have 
\bea\label{appendix-dic1}
G_{xz,xz}= h_{xz}^{(2)} +\Big(4 f_3-\frac{7M^4}{12}+2MO-\frac{M_b}{3}-\frac{M^4\lambda}{2}\Big) \,,~~~~
G_{xz,yz}= h_{yz}^{(2)} 
\eea
with source for $h_{xz}$ and sourceless condition for $h_{yz}$
while 
\bea
\label{appendix-dic1b}
G_{yz,yz}=  h_{yz}^{(2)} +\Big(4 f_3-\frac{7M^4}{12}+2MO-\frac{M_b}{3}-\frac{M^4\lambda}{2}\Big)\,,~~~~
G_{yz,xz}= h_{xz}^{(2)}
\eea
with source in $h_{yz}$ and  sourceless condition for $h_{xz}$. 

\subsection{Results for longitudinal viscosities}

Thus from (\ref{appendix-dic1}), (\ref{appendix-dic1b}) and the result we obtained in section \ref{secb1} we have
\be
\text{Im} G_{xz,xz}=\text{Im} G_{yz,yz}=\omega \frac{f_1^2}{\sqrt{h_1}}\,, ~~~
\text{Im} G_{yz,xz}=4\omega \zeta \frac{q^2 A_{z1}\phi_1^2f_1^2}{h_1}=4\omega\zeta \frac{q^2A_z\phi^2 f^2}{h}\bigg{|}_{r=r_0}\,.
\ee
Thus 
\be
\eta_\parallel =\eta_{xz,xz}=\eta_{yz,yz}=\frac{f^2}{\sqrt{h}}\bigg{|}_{r=r_0}\,,~~
\eta_{H_\parallel}=-\eta_{xz,yz}=\eta_{yz,xz}=4\zeta \frac{q^2A_z\phi^2 f^2}{h}\bigg{|}_{r=r_0}\,.
\ee
Using $s=4\pi f\sqrt{h}\big{|}_{r=r_0}$, We have  $\frac{\eta_\parallel}{s}=\frac{f}{4\pi h}\big{|}_{r=r_0}.$ 
Note that the Chern-Simon term of the gauge fields does not contribute to the formulaes above. This indicates that only the mixed anomaly contributes to the Hall viscosity in the holographic model.

\section{Transverse viscosity}
\label{appc}

To calculate the transverse viscosities we consider the perturbations $\delta g_{xx}=h_{xx}(r) e^{-i\omega t}, \delta g_{xy}=h_{xy}(r) e^{-i\omega t}, \delta g_{yy}=h_{yy}(r) e^{-i\omega t}$ on the background solutions.  Define 
\be\label{def-hL}
h_T=\frac{1}{2}(h_{xx}-h_{yy})\,,~~~Z_\pm=g^{xx}\big(h_T\pm i h_{xy}\big)=\frac{1}{f}\big(h_T\pm i h_{xy}\big),
\ee 
we have
\bea\label{eomlv}
\Big(1\pm\frac{4\zeta \omega}{\sqrt{h}} C_2\Big)Z_\pm'' +\Big[P_2+\frac{2f'}{f}\pm\frac{4\zeta \omega}{\sqrt{h}}\Big(D_2+\frac{2f'}{f}C_2\Big)\Big]Z_\pm'~~~~&&\nonumber\\+\Big[Q_2+\frac{f'}{f}P_2+\frac{f''}{f}\pm\frac{4\zeta \omega}{\sqrt{h}}\Big(E_2+\frac{f'}{f}D_2+\frac{f''}{f}C_2\Big)\Big] Z_\pm&=&0\,,
\eea
where $C_2, P_2, D_2, Q_2, E_2$ are 
\bea\label{app-co2}
C_2(r)&=&2 A_z',\nn\\
P_2(r)&=&\frac{u'}{u}+\frac{h'}{2h}-\frac{f'}{f},\nonumber\\
D_2(r)&=&2 A_z''+2 A_z'\Big(\frac{u'}{u}-\frac{f'}{f}\Big),\\
Q_2(r)&=&\frac{\omega^2}{u^2}-\frac{f'}{f}\big(\frac{h'}{2h}+\frac{u'}{u}\big)+\frac{f'^2}{f^2}-\frac{f''}{f},\nonumber\\
E_2(r)&=&-\Big(\frac{u'}{u}+\frac{f'}{f}\Big)A_z''+\Big(\frac{2\omega^2}{u^2}+\frac{2f'^2}{f^2}-\frac{2f'u'}{fu}-\frac{f''}{f}-\frac{u''}{u}\Big)A_z'\nonumber.
\eea

\subsection{Finite temeprature solutions}
\label{appc1}
At finite temperature and small frequency we can expand $Z_\pm=u^{-i\omega/(4\pi T)}\Big(Z_\pm^{(0)}+\omega Z_\pm^{(1)}+\dots\Big)$, and we have\footnote{The $\omega^2$ term in $E_2$ should be ignored. }
\bea
\Big[\big(uf\sqrt{h}\big)Z_\pm^{(0)'}\Big]'&=&0\,,\nonumber\\
Z_\pm^{(1)''}+\Big(\frac{u'}{u}+\frac{h'}{2h}+\frac{f'}{f}\Big)Z_\pm^{(1)'}\pm\frac{4\zeta }{\sqrt{h}}C_2Z_\pm^{(0)''}+\Big[-\frac{iu'}{2\pi T u}\pm\frac{4\zeta}{\sqrt{h}}\big(D_2+2C_2\frac{f'}{f}\big)\Big]Z_\pm^{(0)'}~~&&\nonumber\\
\Big[-\frac{i}{4\pi T}\Big(\frac{u''}{u}+\frac{u'}{u}\big(\frac{h'}{2h}+\frac{f'}{f}\big)\Big)\pm\frac{4\zeta }{\sqrt{h}}\Big(E_2+\frac{f'}{f}D_2+\frac{f''}{f}C_2\Big)\Big]
Z_\pm^{(0)}&=&0\,.\nonumber
\eea

Thus $Z_\pm^{(0)}=1$ and 
\be
Z_\pm^{(1)}=\frac{i}{4\pi T}\ln u+\int_{r_0}^r\Big[\Big(-if_1\sqrt{h_1}\mp 4\zeta (4\pi T f_1A_{z2})\Big)\frac{1}{uf\sqrt{h}}\pm
4\zeta \Big(\frac{u}{f}\Big)'\frac{f}{u\sqrt{h}}A_z'\Big]d\tilde{r}
\ee with regular boundary condition at the horizon and sourceless boundary condition at the boundary, where $f_1, h_1, A_{z2}$ are horizon values of $f, h, A_z'.$

Thus we have the following solution at the conformal boundary (using the relation in footnote \ref{foot11})
\bea
h_{T}\pm i h_{xy}&=&f u^{-\frac{i\omega}{4\pi T}}\Big(1+\frac{i\omega}{4\pi T}\ln u+\omega\int_{r_0}^r\Big[\Big(-if_1\sqrt{h_1}\mp 4\zeta(4\pi T f_1A_{z2})\Big)\frac{1}{uf\sqrt{h}}\nn\\
&&~~~~~~~~~~~
\pm
4\zeta \Big(\frac{u}{f}\Big)'\frac{f}{u\sqrt{h}}A_z'\Big]d\tilde{r}+\dots\Big)\nonumber\\
&=& r^2-\frac{M^2}{3}+\frac{M^4(2+3\lambda)}{18}\frac{\ln r}{r^2}
+\frac{1}{ r^2}\bigg(f_3+\frac{\omega}{4}\Big(i f_1\sqrt{h_1}\pm 8 \zeta q^2\phi_1^2f_1A_{z1}\Big)\bigg)+\dots
\eea

Similar to the longitudinal case, depending on whether we set $h_{T}$ or $h_{xy}$ to be sourceless, up to first order in $\omega$ we have 
\be
h_{T}= r^2-\frac{M^2}{3}+\frac{M^4(2+3\lambda)}{18}\frac{\ln r}{r^2}
+\frac{1}{ r^2}\Big(f_3+\frac{i\omega}{4}f_1\sqrt{h_1}\Big)+\dots\,, ~~
h_{xy}= -\frac{i\omega}{4 r^2}\Big(8 \zeta q^2\phi_1^2f_1A_{z1}\Big)+\dots\nn
\ee 
or
\be
h_{T}=\frac{i\omega}{4 r^2}\Big(8 \zeta q^2\phi_1^2f_1A_{z1}\Big)+\dots\,,~~~
h_{xy}=r^2-\frac{M^2}{3}+\frac{M^4(2+3\lambda)}{18}\frac{\ln r}{r^2}
+\frac{1}{ r^2}\Big(f_3+\frac{i\omega}{4}f_1\sqrt{h_1}\Big)+\dots\,. \nn
\ee

\subsection{Holographic renormalisation}
Near the conformal boundary, the fluctuations for the transverse viscocities are 
\bea
h_{T}&\simeq&h_{T}^{(0)}r^2+h_{T}^{(0)}\big(-\frac{M^2}{3}+\frac{\omega^2}{4}\big)+
 \frac{h_{T}^{(0)}}{144}\frac{\ln r}{r^2}\big(16 M^4+24 M^4\lambda-6 M^2\omega^2+9\omega^4\big)+\frac{h_{T}^{(2)}}{4r^2}+\dots\,, 
\nn\\
h_{xy}&\simeq&h_{xy}^{(0)}r^2+h_{xy}^{(0)}\big(-\frac{M^2}{3}+\frac{\omega^2}{4}\big)+
\frac{h_{xy}^{(0)}}{144}\frac{\ln r}{r^2}\big(16 M^4+24 M^4\lambda-6 M^2\omega^2+9\omega^4\big)+\frac{h_{xy}^{(2)}}{4r^2}+\dots\,. \nn
\eea

Up to the quadratic order in perturbations, we have the following renormalized on shell action
\bea
S_{\text{on-shell}}&=&
\int \frac{d\omega}{2\pi} d^3x \bigg[ h_T^{(0)}(-\omega)h_T^{(2)}(\omega)+h_{xy}^{(0)}(-\omega)h_{xy}^{(2)}(\omega)-8i\alpha \omega^3 bh_T^{(0)}(-\omega)h_{xy}^{(0)}(\omega)
\nonumber\\&&~~~~
+8i\alpha \omega^3  b h_{xy}^{(0)}(-\omega)h_T^{(0)}(\omega)
+h_{xz}^{(0)}(-\omega)h_{xz}^{(2)}(\omega)+h_{yz}^{(0)}(-\omega)h_{yz}^{(2)}(\omega)
+\text{contact terms}\bigg],\nonumber
\eea
where 
\bea
\text{contact terms}&=& 
\big(h_{T}^{(0)}(-\omega)h_{T}^{(0)}(\omega)
+h_{xy}^{(0)}(-\omega)h_{xy}^{(0)}(\omega)\big)\Big(-8 f_3+\frac{7M^4}{36}\nn\\&&
~~~~~-2MO-\frac{M_b}{3}+\frac{M^2\omega^2}{8}-\frac{3\omega^4}{16}\Big)\,.
\eea
Thus if we normalize the source terms to be 1, up to the first order in $\omega$ we have 
\bea\label{appendix-dic2}
G_{xy,xy}= h_{xy}^{(2)} +\Big(-8 f_3+\frac{7M^4}{36}-2MO-\frac{M_b}{3}\Big)\,,~~~~
G_{xy,T}= h_{T}^{(2)}
\eea
with source in $h_{xy}$ and sourceless boundary condition for $h_{T}$ while
\bea
\label{appendix-dic2b}
G_{T,T}=  h_{T}^{(2)} +\Big(-8 f_3+\frac{7M^4}{36}-2MO-\frac{M_b}{3}\Big)\,,~~~~
G_{T,xy}= h_{xy}^{(2)}
\eea
with source in $h_{T}$ and sourceless boundary condition for $h_{xy}$. 

\subsection{Results for transverse viscosities}
From (\ref{appendix-dic2}) and the solution found in section \ref{appc1} we have
\be
\text{Im} G_{xy,xy}=\text{Im} G_{T,T}= \omega f_1\sqrt{h_1}\,, ~~~
\text{Im} G_{xy,T}=-\text{Im} G_{T,xy}=\omega 8\zeta q^2\phi^2fA_z\Big{|}_{r=r_0}\,.
\ee
Thus we have
\be \eta_{\perp}=\eta_{xy,xy}=\eta_{T,T}=f\sqrt{h}\Big{|}_{r=r_0},~~ \eta_{H_{\perp}}=\eta_{xy,T}=-\eta_{T,xy}=8\zeta q^2\phi^2fA_z\Big{|}_{r=r_0},\ee
and $\frac{\eta_\perp}{s}=\frac{1}{4\pi}.$

\section{On the spin 1 sector}
\label{appd}
For completion, we also study the other modes of spin 1 sector of this system which has not been studied so far. Consider fluctuations $\delta g_{tx}=h_{tx} e^{-i\omega t},~\delta g_{ty}=h_{ty} e^{-i\omega t}$ on the background solutions and they are responsible for the thermal conductivities of the dual system.
Define $s_\pm=h_{tx}\pm ih_{ty}$, and we have the following equations
\be
\Big(1\mp\frac{4\zeta \omega}{\sqrt{h}}A_z'\Big)\big(s_\pm'-\frac{f'}{f}s_\pm\big)=0.
\ee This equation reduces to $s_\pm'-\frac{f'}{f}s_\pm=0.$ The solution to this equation is $s_\pm=c f$, where $c$ is a  constant. Thus the resulting thermal conductivities are $\kappa_{x}=\kappa_{y}=0$ as expected since the underlying  system is at zero density.

\end{document}